\documentclass[final,3p, times, twocolumn]{elsarticle}
\pdfoutput=1

\usepackage{amssymb}

\biboptions{sort&compress}

\journal{Journal of Power Sources}

\begin{document}

\begin{frontmatter}



\title{Current-driven flow instabilities in large-scale liquid metal batteries, and how to tame them}


\author{Norbert Weber\corref{cora}}
\author{Vladimir Galindo}
\author{Frank Stefani}
\author{Tom Weier}
\cortext[cora]{email:~norbert.weber@hzdr.de,
Tel.:~++49\,351\,260\,3112,\\
Fax:~++49\,351\,260\,2007}

\address{Helmholtz-Zentrum Dresden -- Rossendorf, %
Bautzner Landstr.\ 400, %
D-01328 Dresden, %
Germany}

\begin{abstract}
The use of liquid metal batteries is considered as
  one promising option for electric grid stabilisation. While large 
  versions of such batteries are preferred in view of the 
  economies of scale, 
  they are susceptible to various magnetohydrodynamic instabilities 
  which  imply a risk of short-circuiting the battery due to the
  triggered fluid flow. Here we focus on the current driven Tayler 
  instability and give critical electrical
  currents for its onset as well as numerical estimates for the 
  appearing flow structures and speeds. Scaling laws for
  different materials, battery sizes and geometries are found.
  We further discuss and compare various means for preventing 
  the instability.
\end{abstract}


\end{frontmatter}


\section{Introduction}
\label{introduction}
With the large-scale deployment of renewable energy sources, massive and 
cheap electricity storage becomes indispensable since the major part of
renewable electricity generation (solar, wind) is inherently
fluctuating. Storage is thus essential to balance supply and demand and
to stabilize the power grid.

Given that the potential for pumped storage hydropower is largely exhausted,
electrolytically generated hydrogen, partly processed to
synthesized hydrocarbons, seems to be the only viable option
for long-term large-scale storage on the TWh scale. 
However, the total
efficiency of the conversion chain is relatively low due to the
multitude of process steps involved. Electrochemical energy storage
(EES) shows generally higher efficiencies, but needs improvements
towards larger capacities at significantly 
lower costs \cite{WadiaAlbertusSrinivasan:2011}. If these demands can be met,
EES will be an attractive candidate for short-term and mid-term
stationary electricity storage.

Liquid metal batteries (LMBs) are currently discussed as a means to
provide economic grid-scale energy storage
\cite{Bradwell2012}. Typically, an LMB consists of a liquid alkaline
metal layered atop a molten salt which itself floats on a molten metal
or half metal. On discharge, the alkaline metal of the anode is
oxidized and cations enter the fused salt which is often an eutectic
composed of alkali halides. At the electrolyte/cathode interface,
alkaline ions leave the electrolyte, are reduced and alloy with the
cathodic metal. For this setup to work, the alkaline metal needs to be
lighter than the metal forming the cathode, and the molten salt's
density must be in between those of both metals. A battery with a
fully liquid active interior has a number of advantages: The battery
is self-assembling due to stable density stratification. Liquid-liquid
interface processes possess fast kinetics, thereby allowing for fast
charging and discharging and high current densities (about 4 -
100\,kA\,m$^{-2}$). Structureless electrodes are insusceptible to
aging which allows for potentially unlimited cyclability. Scale-up on
the cell level is facilitated by the simple cell
construction. Drawbacks include the elevated operation temperature and
the relatively low cell voltage (typically below 1\,V).

LMBs, or molten salt batteries \citep{Swinkels1971}, were intensively
investigated in the 1960s mainly as part of energy conversion systems,
i.e., thermally regenerative electrochemical systems (TRES).
Interestingly, already at that time the use of LMBs for off-peak
energy storage received interest as well as funding
\citep{Steunenberg2000}. Among the systems investigated early were
Na-Sn \cite{Weaver1962,Agruss1963} and K-Hg \cite{Henderson1963}
because they can be easily separated by distillation. In the quest for
higher cell voltages and current densities, among others, Na-Bi,
Li-Te, and Li-Se cells were built, abstaining from thermal
separability for the two latter systems.  Progress in the field has
been reviewed several times in the 1960s and early 1970s: Crouthamel
and Recht \citep{Crouthamel1967} edited a special volume of {\it
  Advances in Chemistry} in 1967 dedicated to ``regenerative EMF
cells''. The large body of research performed at Argonne National
Laboratory was discussed in detail by Cairns et al.\ \cite{Cairns1967}
and described in a more compressed version by Cairns and Shimotake
\cite{Cairns1969b}.  Swinkels' \citep{Swinkels1971} discussion had a
slightly broader scope.  A very recent account is due to Kim et al.\
\cite{Kim2013b} who not only review the previous achievements, but
report on the ongoing research at the Massachusetts Institute of
Technology (MIT) as well. The focus of the MIT activities is on the
deployment of LMBs for large-scale stationary energy storage
\cite{Bradwell2012}.  Consequently, emphasis is now on different
practical and economical aspects as the utilization of abundant and
cheap active materials \cite{Kim2013a}.

\begin{figure}[hbt]
\centerline{
\includegraphics[width=\columnwidth]{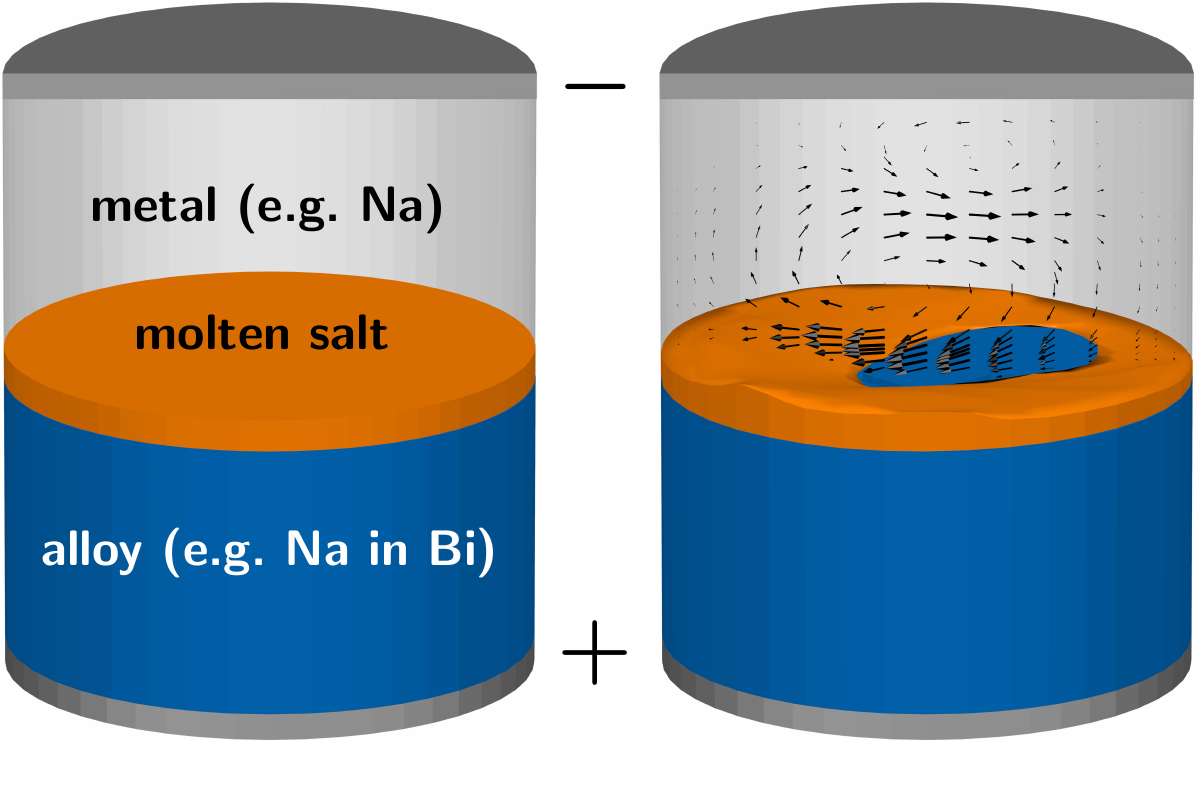}}
\caption{Sketch of a liquid metal battery with typical inventory (left). 
The electrolyte works as ion conductor and separates the two liquid metals. 
A movement of the fluid, as it may result from magneto-hydrodynamic
instabilities, could wipe the electrolyte and lead to a battery 
failure (right).}
\label{fig:battery}
\end{figure}

Besides the chemistry of different material combinations, and the
related corrosion problems, there are further aspects of LMBs which
should not be overlooked.  For traditional battery systems (apart from
lead-acid and redox flow batteries) fluid dynamics does not play a
significant role. This is different for LMBs due to their completely
liquid interior. On the one hand, mass transfer in the electrolyte and
the cathodic compartment can be facilitated by liquid motion. On the
other hand, the density stratification -- appearing stable at first
glance -- might be disturbed by flow instabilities (see Fig.\ 
\ref{fig:battery}). Here we will focus on a special
magneto-hydrodynamic phenomenon. After a few short remarks on
interface instabilities in current carrying systems, we will proceed
to the current-driven, kink-type Tayler instability, which is the main
subject of the paper. It should be noticed that the significance of
hydrodynamically stable interfaces for the operation of LMBs was
recognized early, see, e.g., Cairns et al.\ \cite{Cairns1967} and
Swinkels \cite{Swinkels1971}.

Interface instabilities are well known phenomena in Hall-H\'eroult,
i.e.,
aluminum reduction cells (ARCs)
\citep{Sele1977,BojarevicsRomerio1994,Davidson2000,Evans2007,Molokov2011}.
These instabilities result from the interaction of the ionic current
with a background magnetic field generated by the current supply lines.
As a consequence, waves with a length comparable to the cell width
develop and culminate in a sloshing motion of the aluminum. Amplitudes
may become high enough for the aluminum to contact the graphite anodes,
effectively short-circuiting the cell and terminating the reduction
process. To keep the long-wave instabilities at bay, the cell current
has to be kept below a critical value and the cryolite layer should not
be thinner than about 4.5\,cm. These requirements lead to a total cell
voltage of about 4.5\,V. 2\,V of those, i.e., around 40\% of the cell
voltage are spent on overcoming the ohmic resistance of the electrolyte
layer, the corresponding power is converted to
heat \citep{Davidson2000}. Since expenses for electricity are a major
part of the overall costs of aluminum production, a reduction of the
electrolyte thickness even by a few millimeters only would result in
huge savings, but is impeded by the long-wave interface instabilities
described above.

ARCs have two liquid layers sandwiched between the carbon anode on top
and the current collector on the bottom. In contrast, LMBs are three
layer systems. Presently, only little is known about current driven
interface instabilities in such settings. Sneyd \citep{Sneyd1985}
predicts theoretically that additional short wave instabilities may
arise in the case of the two neighboring liquid-liquid interfaces. To
the authors' best knowledge, no experimental observations of current
driven instabilities in three layer systems with closely spaced
interfaces have been reported to date. Hoopes cells, i.e., three-layer
cells used for the electrolytic refining of aluminum, apparently do not
suffer from such instabilities. The electrolyte layer of these cells is
quite massive with about 8 to 10\,cm thickness
\cite{Dube1954,Srinivasan1986}. Thus, a strong
interaction of both interfaces is relatively unlikely. However, violent
swirling motions driven by non-uniform electric and magnetic fields
emerging from the cathodic current collectors were observed already
quite early \citep{Edwards1930}. These flows
occurred, if the cathode metal thickness was chosen too small. As a
result, spotty contacts of anode and cathode layers could happen. 

Coming back to LMBs, the maximum thickness 
of the electrolyte layer is limited by the requirement that the voltage 
loss in the electrolyte must not exceed the available open circuit voltage (OCV).
Indeed, a meaningful design requires the voltage losses to be much
smaller than the OCV. For the most relevant combinations of metals and 
salts, this means the electrolyte's thickness should 
not exceed a few millimeters. 
Under those conditions, a careful consideration of the 
stability of the interfaces becomes imperative.

Exploiting the economies of scale is the usual route towards economic
devices. Since current densities are large, cell scale-up (up to a few
cubic meters \cite{Sadoway2012}) will result in considerable total
cell currents. These large currents can trigger the so-called Tayler
instability (TI), as has been shown in a recent liquid metal
experiment by Seilmayer et al.\ \cite{Seilmayer2012}. The TI is a kink
type instability very similar to that occurring in the z-pinch of
plasma physics. TI manifests itself in one or, depending on the cell's
aspect ratio, more vortices rotating around an axis normal to that of
the cell. Weber et al.\ \cite{Weber2013} developed an
integro-differential equation solver (in the sense of Meir et al.\
\cite{Meir2004}) capable of reproducing the experimental findings in
minute detail.  Building upon these results, we will use the numerical
toolbox to discuss the consequences of the TI for the construction of
larger LMB cells. Starting with a discussion of the proper
dimensionless number describing the TI's onset in different materials,
we will proceed to the influence of the cell's aspect ratio on the TI,
and to various possible countermeasures. The first of these measures
consists in modifying the cell interior with a central bore.  Second,
we will feed a current through a conductor placed in this central
bore. Both methods were derived from stability theory by Stefani et
al.\ \cite{Stefani2011} and are here numerically confirmed to be
effective. Third, a stabilization method based on imposing an axial or
horizontal magnetic field, which requires no modification of the
cell's interior, will be discussed. Obviously, such a method would be
preferred over more invasive approaches. All three stabilization
methods act by modifying the electromagnetic field in the cell.

Mechanical means of flow stabilization, like the use of baffles to
break long-wave motions, are limited by the need to find materials
capable of withstanding the aggressive environment
\cite{Davidson2000}.  Additionally, an immobilization of the
electrolyte by using pastes or ceramic matrices results in
considerably larger values of the internal cell resistance. A six-fold
increase has been found by Heredy et al.\ \cite{Heredy1967}, similar
values can be deduced from Cairns and Shimotake \cite{Cairns1969b}.

Finally, we would like to reiterate that not every fluid motion 
is necessarily dangerous for LMBs. Quite in contrast, 
under certain circumstance it can even be beneficial, in particular 
in the cathodic compartment. There, mass transport limitations and
associated problems like concentration polarization and local
solidification at the alloy-electrode interface can be mitigated or
avoided by agitation. Diffusion control in bimetallic cells was observed
by several researchers
\cite{Agruss1967,Heredy1967,Bradwell2012,Kim2013b}
as well as the formation of intermetallics
\cite{Shimotake1967,Kim2013a}. Equally early it
was obvious that stirring would be a remedy
\cite{Foster1967b,Shimotake1967,Swinkels1971}. Giving the high
temperature environment, electromagnetic agitation constitutes an
attractive means requiring no direct contact with the hot metal or
construction material. Yet, even when considering the possible
advantages of flow instabilities for LMBs, they have to be 
well understood in the first instance. This is the  main goal of 
the present paper.

\section{The Tayler Instability in liquid metal batteries}
Typically, liquid metal batteries are assumed to 
have the shape of a cylinder or a cuboid. 
Their aspect ratio, i.e.\ the ratio of height to 
horizontal extension, depends on the envisioned 
storage capacity. 
Flat batteries have low capacities, but, as we will see below,  
are less vulnerable to the TI than taller batteries
with higher capacity. While the economies of scale 
demand for liquid metal batteries with a large size in general, 
thermal insulation aspects make a point for
choosing rather compact design with an aspect ratio 
close to one. 

Postponing this aspect ratio issue to the 
next section, here we would like to start with  
characterizing the TI in a paradigmatic geometry.
Focusing on its dependence 
on various material parameters of the liquids to 
be utilized,  
we consider four typical anode materials for which we
will compute the critical currents, the growth rates,
and the velocities in the saturated state of the TI. Note that,
as a rule, cathode materials are less conductive and more dense,
which makes them less susceptible to the TI than the anode 
materials, although this has to be verified for each 
particular material combination.

\subsection{Growth rates and critical currents}

\begin{figure}[h!]
\centerline{
\includegraphics[width=\columnwidth]{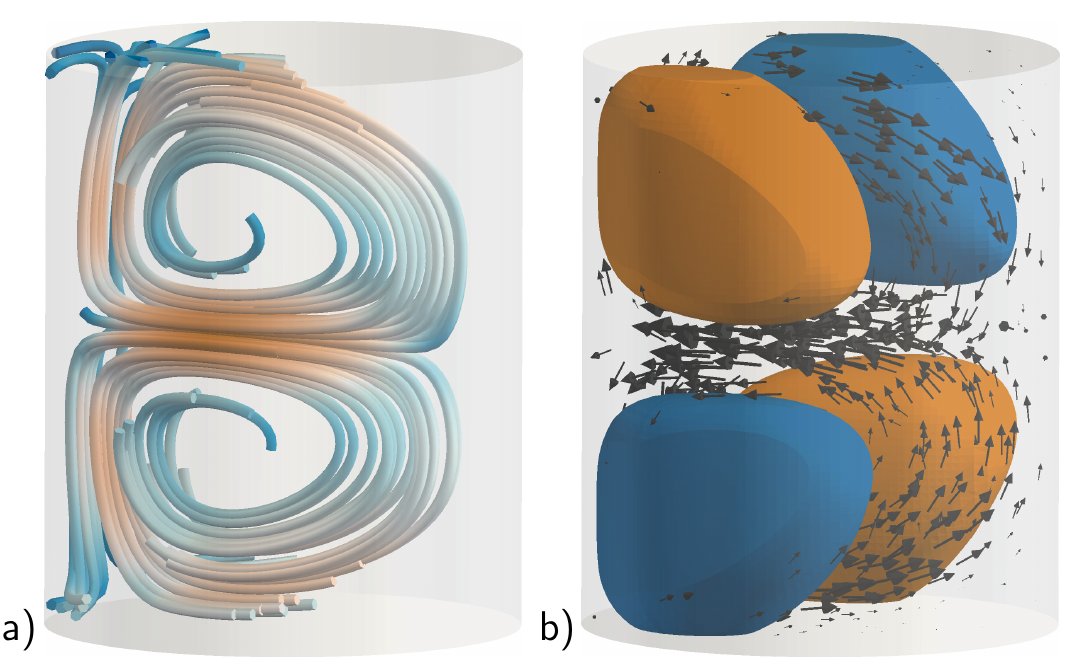}}
\caption{Flow field of the TI (a) in a cylinder 
filled with liquid Na, 
$I = 4$\,kA, material parameters taken at 
$T=365\,^{\circ}$C,  and contours of the corresponding magnetic 
field $b_z$ (b) in the saturated state.}
\label{fig:VzCells}
\end{figure}

Fig.\  \ref{fig:VzCells} illustrates both the flow structure (a) 
and the induced vertical magnetic field component $b_z$ (b) for
an assumed liquid metal column with diameter 
$d=248$\,mm and height $h=300$\,mm, i.e.\ an aspect ratio of
$h/d=1.2$. For this particular 
aspect ratio the TI acquires the form of two vortices,
the highest velocities appearing approximately 
at mid-height of the cell.
The only slightly lower horizontal velocity at the 
bottom of the liquid anode column has to be considered 
with regard to its capability to shear off the thin electrolyte layer 
and hence to destroy the stable stratification of the three layer 
system.

\begin{figure}[h!]
\centerline{
\includegraphics[width=\columnwidth]{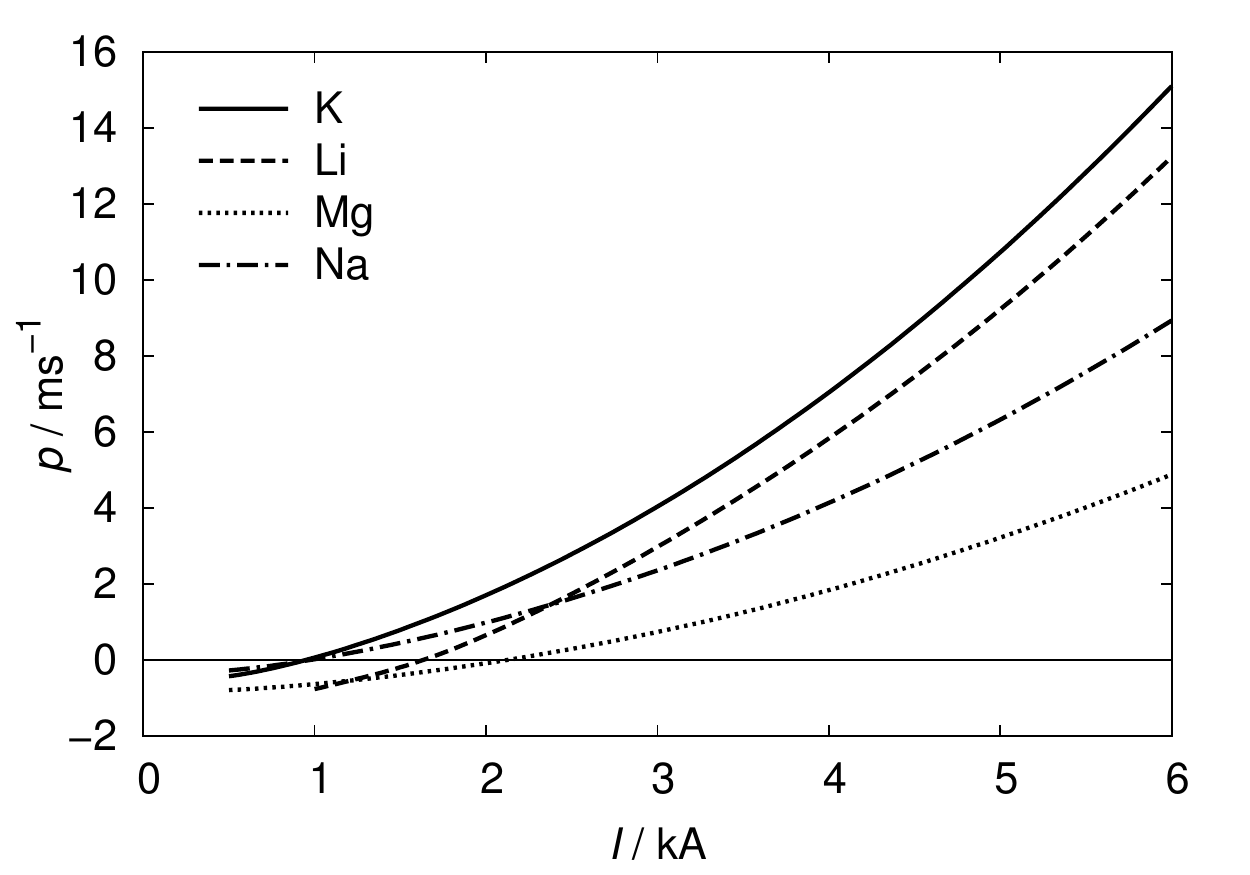}}
\caption{Growth rate of the TI in dependence of the applied current
for different anode materials. The critical current corresponds to a
growth rate of zero.}
\label{fig:gr(I)}
\end{figure}

For this geometry, and the four anode materials K, Li, Na, and Mg,
Fig.\  \ref{fig:gr(I)} shows the
dependence of the growth rate of the TI on the total current
through the column.
The growth rate $p$ of a linear instability 
is defined by the exponential
increase $\sim \exp{p t}$ of its typical perturbations, 
in our case of the perturbations of the velocity and the magnetic field.
The critical current for either material, 
obtained by setting $p=0$, is shown in the second column of 
Table \ref{tab:Icr}.
The third column indicates the corresponding 
Hartmann number, a dimensionless number often used in 
magnetohydrodynamics whose square measures the ratio of magnetic 
forces to viscous forces in a fluid.
Here, the Hartmann number is defined as  
\begin{equation}
{\rm Ha}=B_{\phi}(R) R \sqrt{\frac{\sigma}{\nu \rho}}=
\frac{I}{2 \pi} \sqrt{\frac{\sigma}{\nu \rho}}
\label{eq:hartmann}
\end{equation}
where $B_{\phi}$ is the azimuthal magnetic field at the outer 
radius $R=d/2$, 
$\sigma$ is the electrical conductivity, $\nu$ the kinematic viscosity, and $\rho$ the
density of the fluid. As can be seen in  Eq. (\ref{eq:hartmann}), the
Hartmann number does not depend on the radius of the
battery but only on the total current $I$ through the liquid 

\begin{figure}[h!]
\centerline{
\includegraphics[width=\columnwidth]{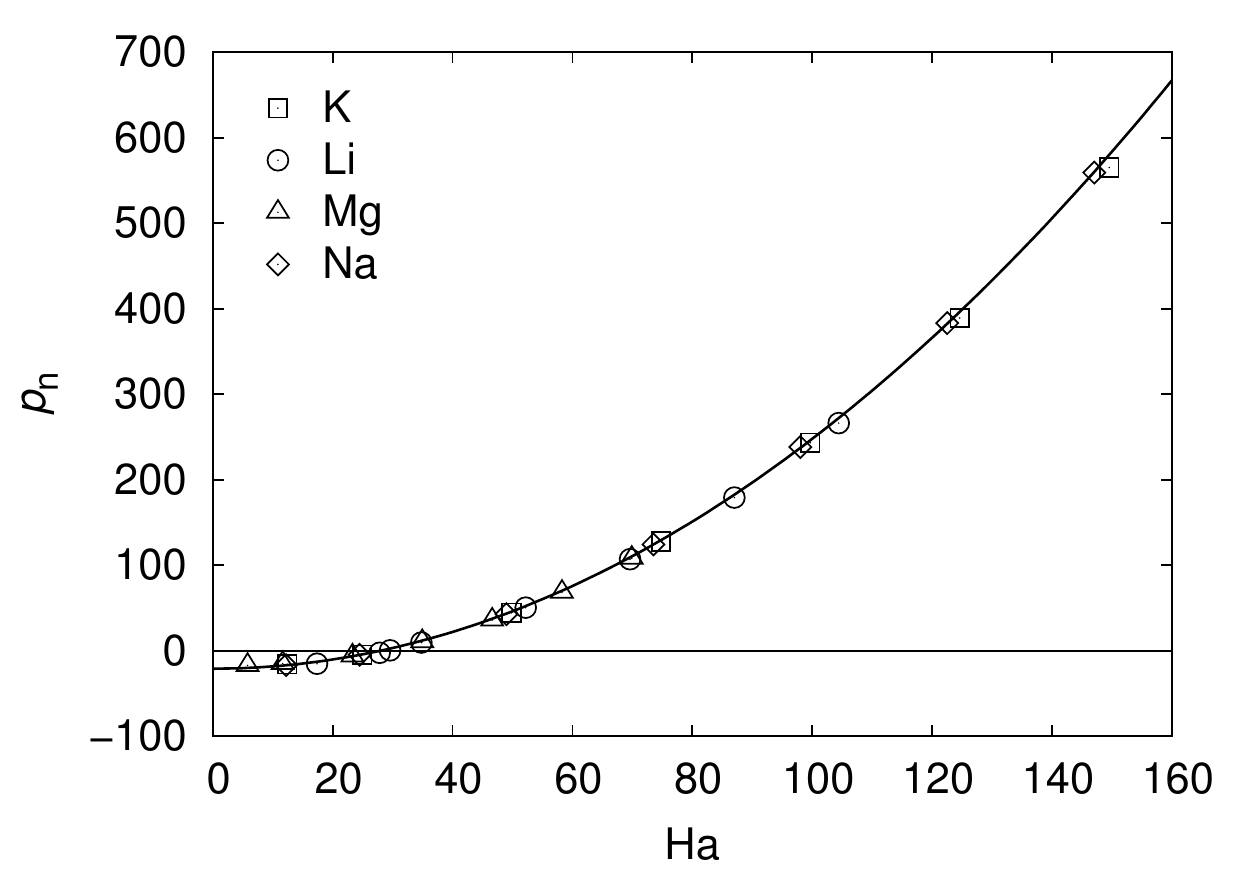}}
\caption{Growth rate of the TI in dependence of the Hartmann number
  for different liquid metals. The growth rate is normalized 
  to $\nu/R^2$.}
\label{fig:gr(Ha)}
\end{figure}

Evidently, we obtain the same critical Hartmann number of 29 
for all the considered materials which means 
that $\rm Ha$ is indeed the governing dimensionless number 
for the onset of the TI. We can further define a dimensionless 
growth rate of the TI by dividing the physical growth rate $p$
by the inverse of the typical viscous time scale $R^2/\nu$. 
Fig.\  \ref{fig:gr(Ha)}
shows then this dimensionless growth rate $p_n = p \cdot R^2/\nu$ 
in dependence on  
$\rm Ha$. Not surprisingly, the curves for all four  
materials collapse into one single curve. This means, in 
turn, that for any concrete anode or cathode material we could
work with the critical $\rm Ha$ (which is 29 for the chosen 
aspect ratio) and derive from that the critical current.

\begin{table}[h!]
\centering
\caption{Critical currents, and critical Hartmann numbers, 
for the onset of the TI in an anode liquid 
metal column with aspect ratio $h/d=1.2$.}
\label{tab:Icr}
\begin{tabular}{lcc}
\hline
\strut{}metal & $I_{cr} /$ A & Ha$_{cr}$\\
\hline
K  & 1163 & 29\\
Li & 1666 & 29\\
Mg & 2488 & 29\\
Na & 1183 & 29\\
\hline
\end{tabular}
\end{table}

\subsection{Saturation level}

\begin{figure}[h!]
\centerline{
\includegraphics[width=\columnwidth]{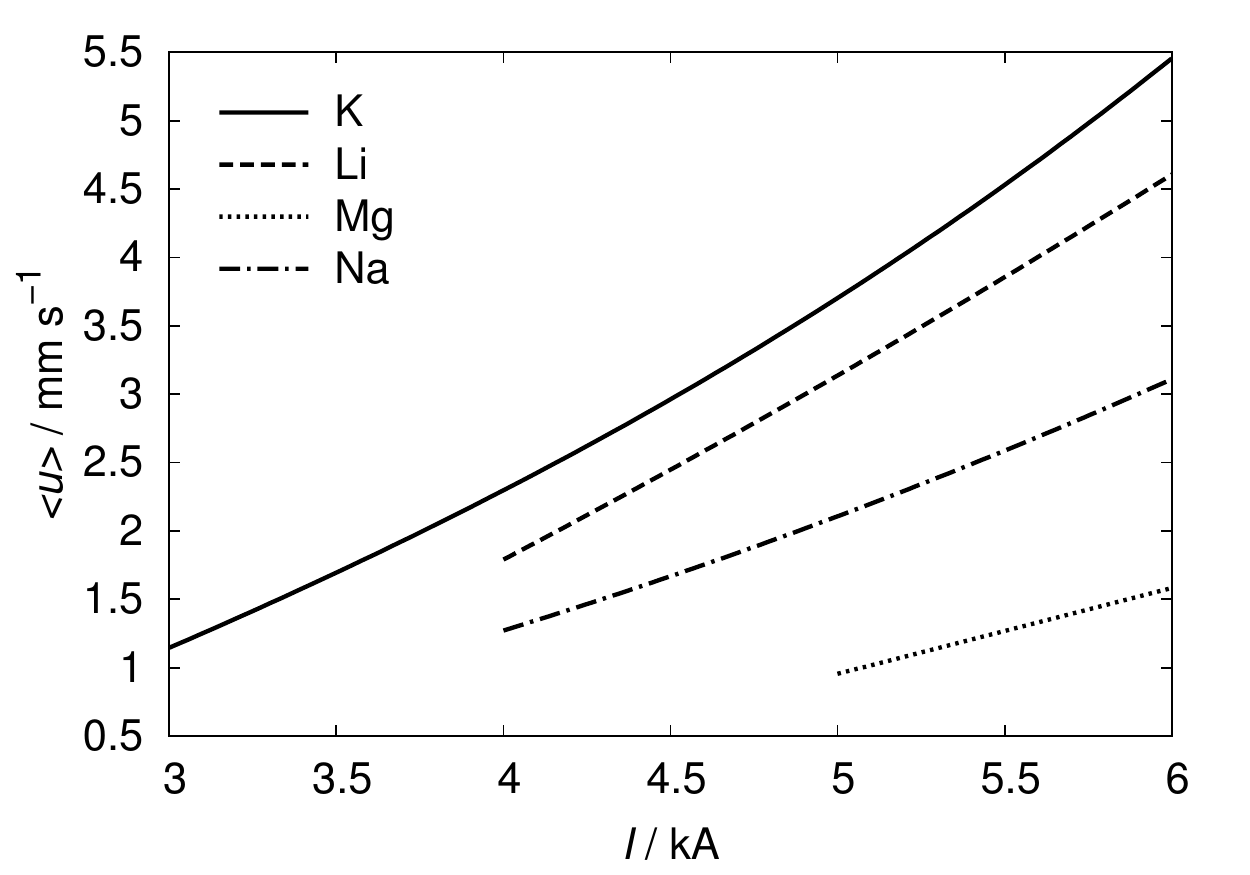}}
\caption{Mean velocity of the saturated TI for different liquid metals
  in dependence of the applied current.}
\label{fig:v(I)}
\end{figure}

A quite analogous procedure can be applied when determining the
dependence of the saturated velocity on the total current.  For each
of the considered liquid metals, in Fig.\ \ref{fig:v(I)} we plot the
rms value of the velocity in dependence on the current. For the
particular case of using K, and a current of 6\,kA, this rms value
reaches a value 5.5\,mm\,s$^{-1}$. The maximum speed at the bottom is
approximately double that value, i.e.\ 1\,cm\,s$^{-1}$, which may
already be considered a danger for the integrity of a thin electrolyte
layer below it.

\begin{figure}[h!]
\centerline{
\includegraphics[width=\columnwidth]{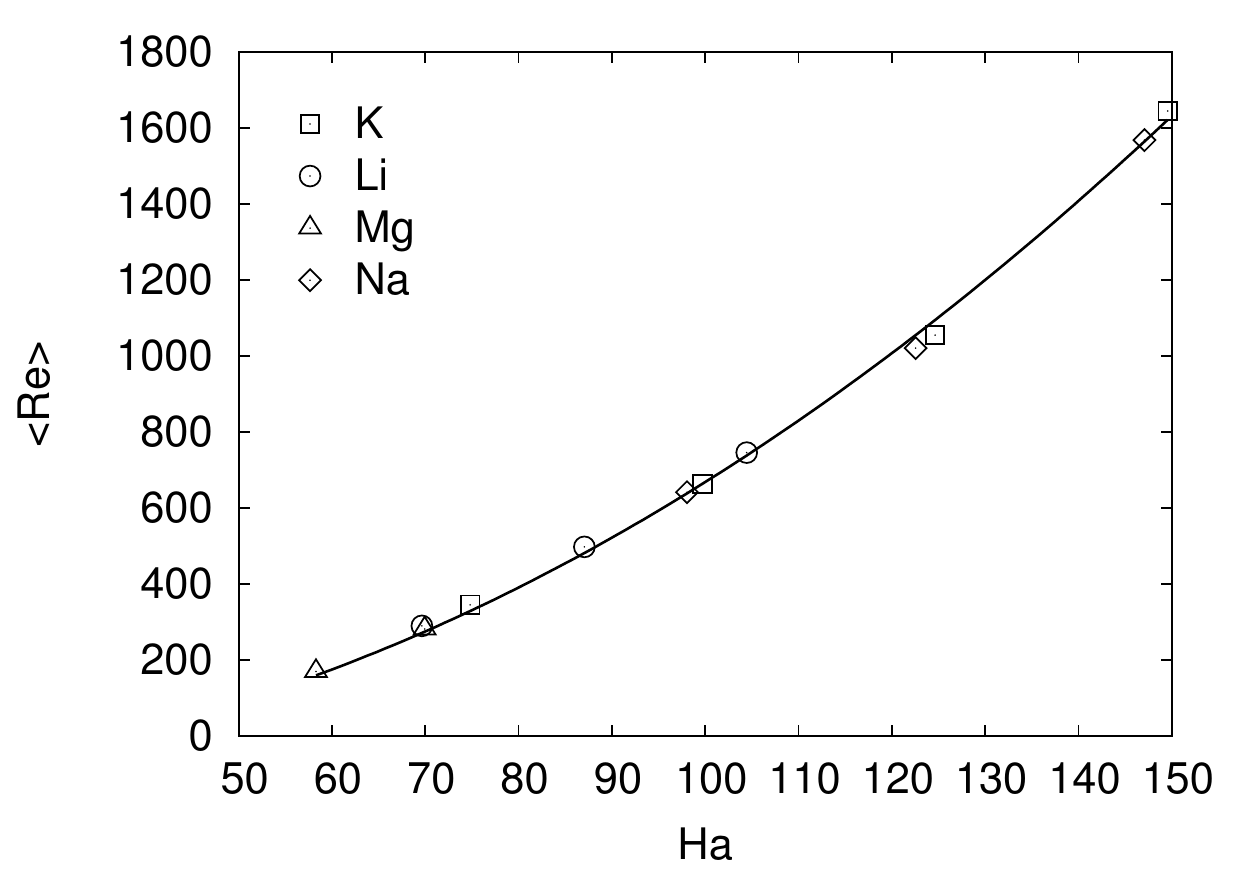}}
\caption{Mean Reynolds number of the saturated TI for different liquid
  metals in dependence of the Hartmann number.}
\label{fig:Re(Ha)}
\end{figure}

Again, these four curves can be collapsed into a single one (see 
Fig.\  \ref{fig:Re(Ha)}) 
when going over from the physical quantities to the 
appropriate dimensionless parameters. On the 
abscissa, this is again $\rm Ha$, while the appropriate 
dimensionless number for measuring the mean velocity $\langle u\rangle$ 
is the Reynolds number, defined by
\begin{equation}
{\rm Re}=\langle u\rangle R/\nu  \; .
\label{eq:reynolds}
\end{equation} 
On closer inspection, and based on previous computations \cite{Ruediger2011}, 
it turns out that for large currents 
$\rm Re$ grows quadratically with
$\rm Ha$, so that for a given electrolyte layer thickness there 
is definitely a current at which the TI would destroy the stable 
stratification.                     	   

To summarize this section, we have seen that the TI is completely 
governed by the Hartmann number, which determines both the 
non-dimensionalized growth rate in the exponential growth phase
as well as the Reynolds number of the saturated flow. For 
any concrete material and size of the
liquid column, Figs.\ \ref{fig:gr(Ha)} and \ref{fig:Re(Ha)}
would allow to determine the physical growth rates 
and the final velocity scale by using the definitions 
(\ref{eq:hartmann}) and (\ref{eq:reynolds}), respectively.

\section{Taming the Tayler Instability}
Having seen that for large charging or discharging currents
the TI may indeed pose a problem for the stability of the 
liquid three-layer system, in this section we will discuss various 
methods how to tame it. The first idea that comes to mind here 
is the insertion of some non-conducting, e.g., ceramic) 
blades into the battery for preventing or at least damping the 
TI triggered flow. As simple as this may sound, 
it is not without problems, a serious one being 
connected with the fluctuating height of the layers during
charging and discharging which could result in thin
metal layers adhering to the blades that would lead to a 
short circuit. 
A similar method could rely on using a swimming "carpet" 
below or above the electrolyte  
in order to stabilize the latter. 
Just for completeness, we also note the somewhat fancy 
possibility to tame the TI by setting 
the total battery into rotation \cite{Ruediger2011}.

While not excluding any viable solution based on inserting 
blades, or felts, into the battery, in the following we will 
exclusively focus 
on those solutions that maintain the advantageous simplicity of 
a compact fluid inventory that is not separated into more 
different segments than necessary. 
We will indeed see that the intrinsic dependence of the TI on geometric 
factors, on one side, and on external magnetic fields, on  the other side,
allows for quite simple methods to suppress it.
Accordingly, we will exploit the effect that the critical current
increases with decreasing height of the 
liquid metal layers, but also with the radius of an
supposed central bore. Second, we will
see that 
a central return current through such a bore 
would completely prevent the TI from occurring, and that
the same stabilization effect can be achieved by applying a
vertical or horizontal magnetic field.

\subsection{Dependence on the aspect ratio}

As many other flow instabilities, the TI is characterized by an
optimal axial wavelength $\lambda$ for which its growth rate and its
saturated velocity scale become maximal.  This means, in turn, that if
the height of the liquid column falls below the half of this optimal
wavelength, the instability is significantly damped or even
suppressed. For that reason, flat batteries are much less vulnerable
to the TI.  Yet, if a high storage capacity per cell is desired, the
liquid metal columns must be tall enough. Assuming a current density
of 10$\,$kA\,m$^{-2}$, a 10\,mm thick sodium layer is transferred per hour
from the anodic to the cathodic compartment. This quickly adds up to a
meter, if storage capacities of a few days are demanded.

\begin{figure}[h!]
\centerline{
\includegraphics[width=\columnwidth]{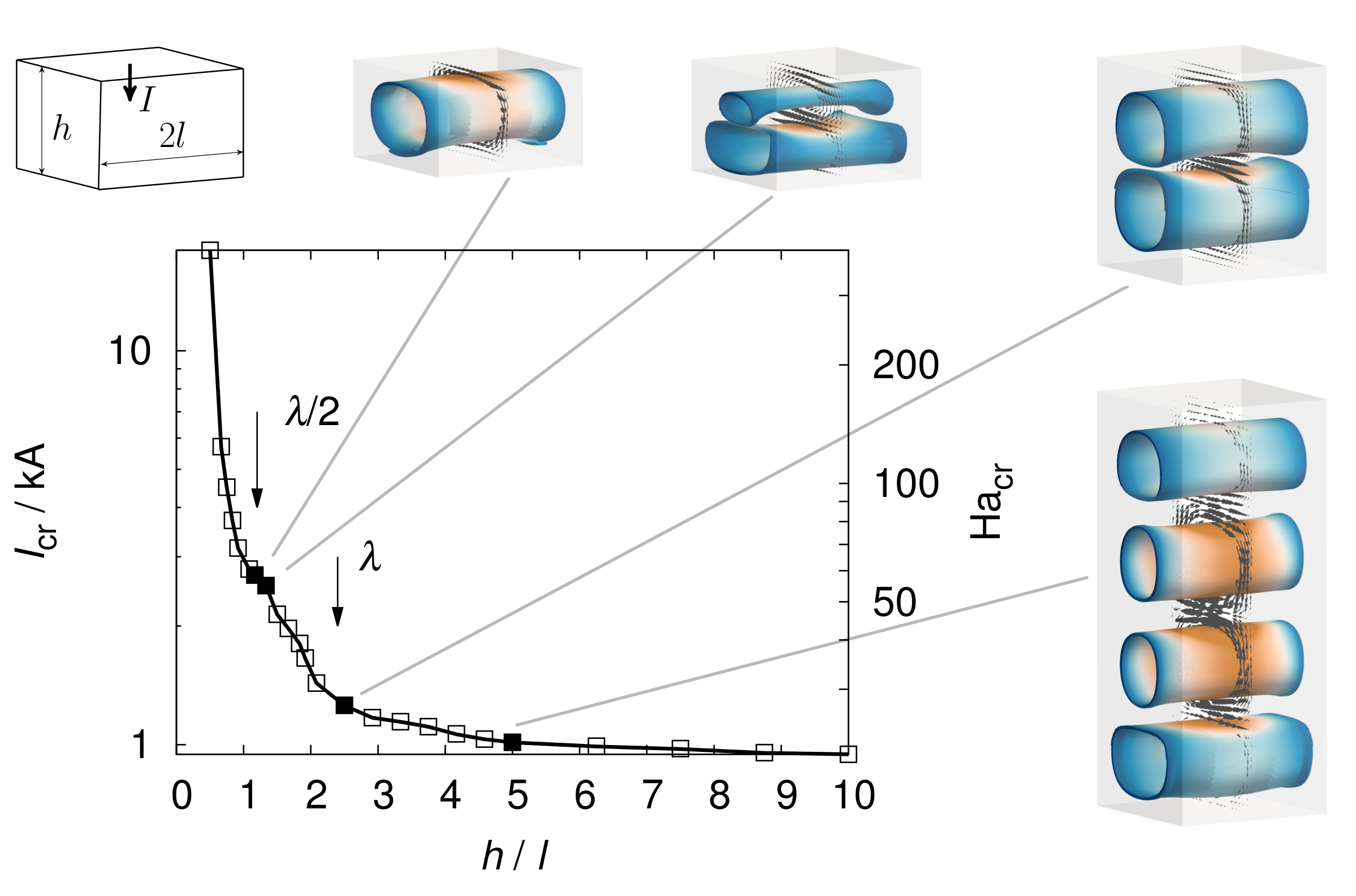}}
\caption{Critical current (Na) and Hartmann number for onset of the TI in dependence of the
  aspect ratio. Simulations were carried out with a cuboid geometry
  with the dimensions $96\times 96\times h$\,mm$^3$.}
\label{fig:aspect}
\end{figure}

Reiterating the results of Weber et al.\ \cite{Weber2013}, in Fig.\ 
\ref{fig:aspect} we illustrate the dependence of the critical Hartmann
number on the aspect ratio $h/l$ of a cuboid with height $h$ and width
$2l$.  Starting at high values of $h/l$, we first observe a very
moderate increase of ${\rm Ha}_{cr}$ which becomes significantly
steeper after having passed a sort of plateau at $h/l \sim 2.4$. This
is the aspect ratio where just one wavelength of the TI fits into the
column.  Interestingly, a second plateau forms approximately at $h/l
\sim 1.2$, where just half a wavelength fits into the column. For even
shallower columns, a very steep increase of ${\rm Ha}_{cr}$ can be
observed.  It should be noticed that this sensitive dependence may
also lead to a transient behaviour of the TI when the height of the
better conducting layer (i.e., in most cases the upper one) changes
during the charging/discharging process.

\subsection{A central bore in the battery}

A simple trick to increase the critical current 
had already been devised in a recent paper \cite{Stefani2011}. 
Assuming 
a column of infinite length, it was shown that the mere 
existence of a central bore 
in the middle of the cylinder would increase the critical 
current. 
Even when taking into account the corresponding area reduction, 
the allowed total current through the remaining area is still 
higher than without this central bore.

\begin{figure}[h!]
\centerline{
\includegraphics[width=\columnwidth]{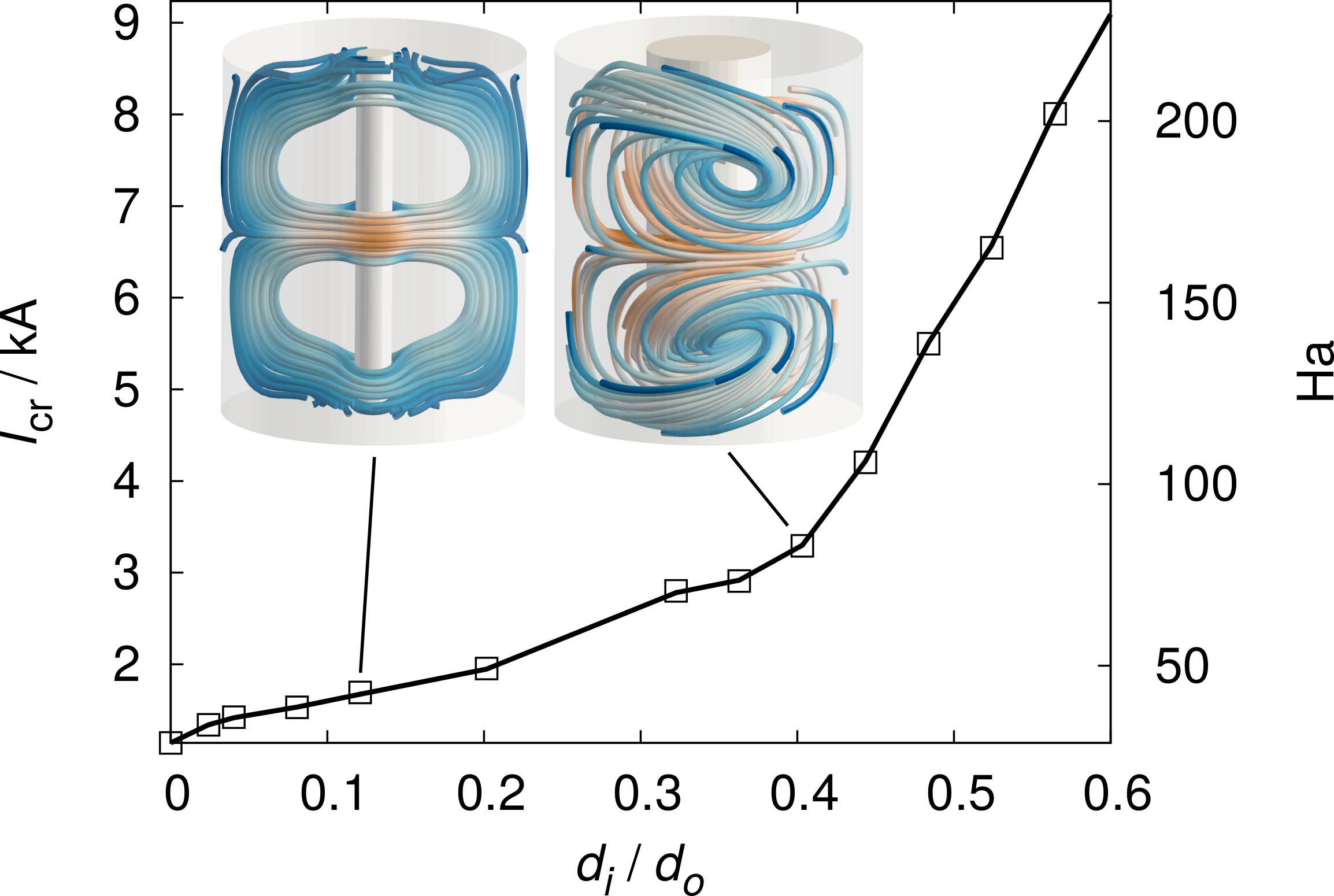}}
\caption{Critical current for an electrode with a bore in the middle
  ($d_o = 248$\,mm, $h = 300$\,mm, Na).}
\label{fig:hz_I}
\end{figure}

Here, we concretize and confirm this behaviour for the more 
realistic case of a finite height cylindrical column. Again, 
we use the example of an
electrode of liquid Na with an outer diameter of 
$d_o = 248$\,mm and a height of $h = 300$\,mm. 
The inner diameter $d_i$ is assumed variable. 
Increasing this diameter $d_i$ of the bore, 
the critical current for the onset of the TI increases 
more or less linearly. When the
inner radius reaches about 40 \% of the outer radius, the critical currents
start to increase even steeper (Fig.\ \ref{fig:hz_I}). This bend 
of the curve is caused by a change in the flow structure of the TI. 
While the flow
maintains its basic vortex structure for small bores, it acquires a more
spiral shape for larger bores. 
 
\begin{figure}[h!]
\centerline{
\includegraphics[width=\columnwidth]{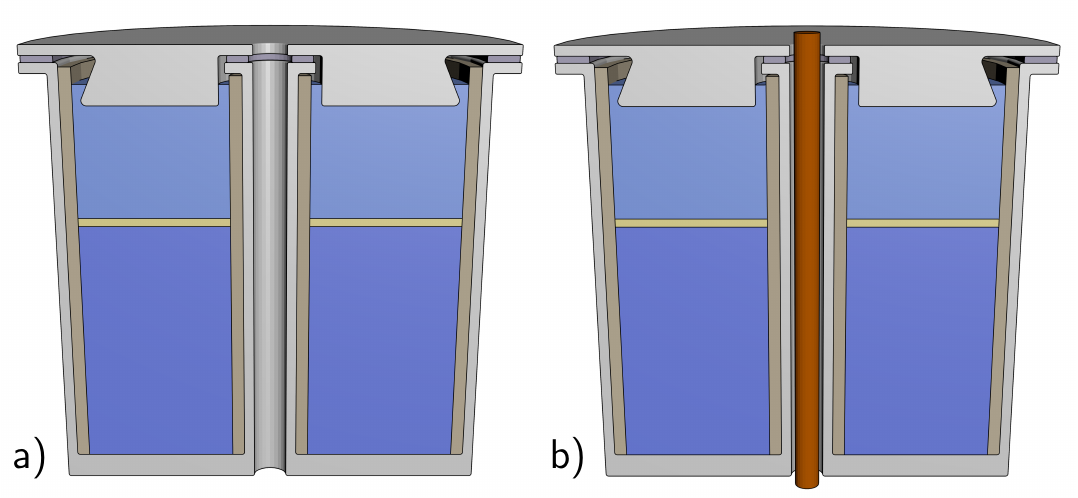}}
\caption{Providing the liquid metal battery with a bore or returning the current
  are effective ways to avoid the TI.}
\label{fig:bore}
\end{figure}

Fig.\  \ref{fig:bore}a illustrates a corresponding technical solution,
with a central bore going through the total battery.
It should be noticed, however, that the gain in terms of
achievable critical current is typically 
only by some factor in the order of 2 -- 10 or so.
A more radical way of preventing the TI will
be discussed in the following. 
 
\subsection{Returning the current}
Imagine now that the central bore is equipped
with a massive wire as indicated in Fig.\  \ref{fig:bore}b.
For the idealized case of an infinite length cylinder, and 
assuming the fluid to be non-viscous and non-resistive, it had been 
shown that the TI
can be completely suppressed by sending an appropriate current
through this wire \cite{Stefani2011}. This additional current 
changes the radial
dependence of the azimuthal magnetic field $B_\phi(r)$ so that 
the ideal stability condition \cite{Tayler1973}
\begin{equation}
\frac{d}{dr}(r B^2_\phi(r))<0
\end{equation}
may be met. To fulfill this condition, one can either use a current 
in the same direction  as the charging/discharging
battery current $I_{bat}$, 
with an amplitude of $I_{wire} \ge I_{bat} (3+(d_i/d_o)^2)/(1-(d_i/d_o)^2)$,
or in the opposite direction with an amplitude $I_{wire} \ge I_{bat}$.
Evidently, the second method needs less current and will
therefore lead to less total losses. It is 
also more convenient from the technical point of view, since the 
charging/discharging current can be easily 
redirected (see Fig.\  \ref{fig:bore}b).

\begin{figure}[h!]
\centerline{
\includegraphics[width=\columnwidth]{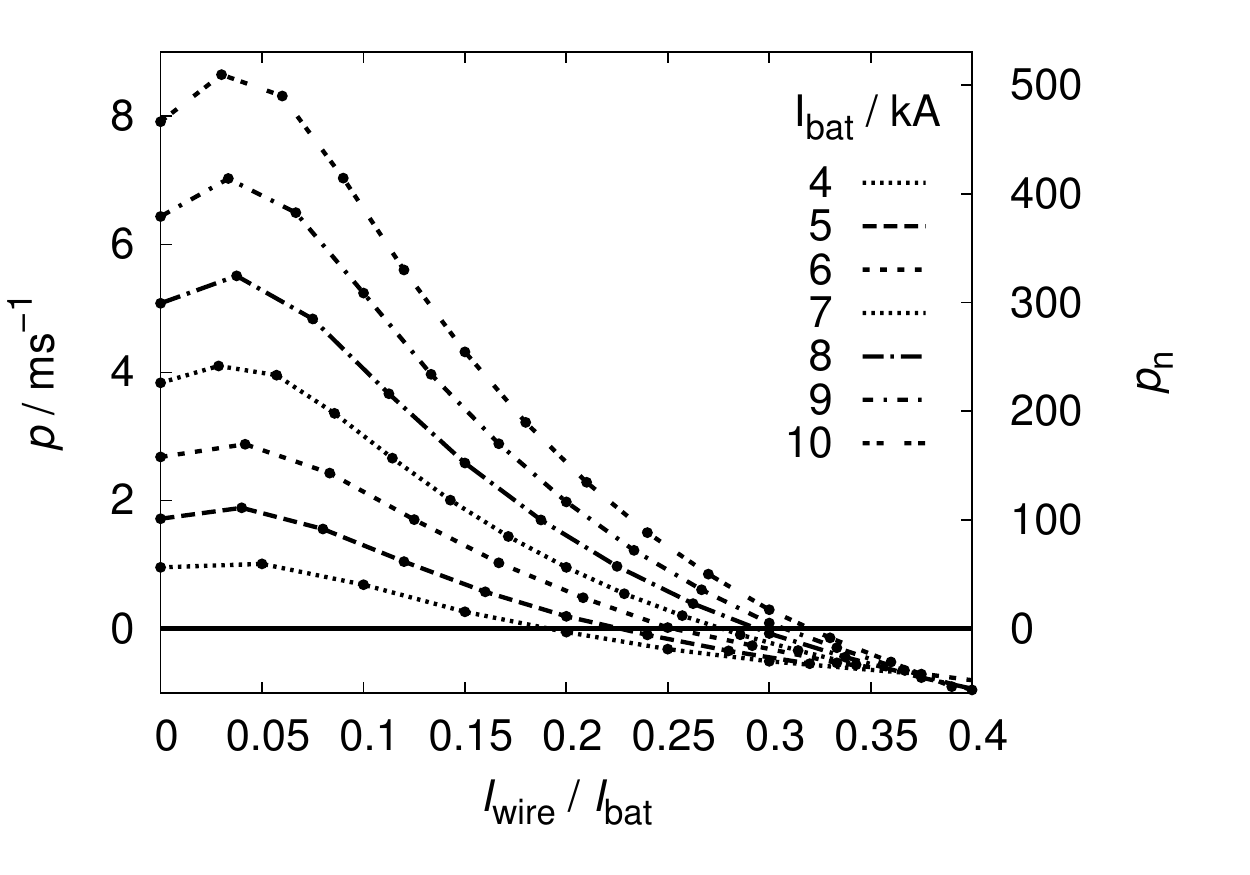}}
\caption{Growth rate of the TI in a liquid metal battery electrode
  with a return current through a bore of $d_i = 30$\,mm ($d_o = 248$\,mm, $h = 300$\,mm, Na).}
\label{fig:rs_gr}
\end{figure}

Here, we consider the more realistic case of a finite height 
column, and we
take into account the real viscosity and resistivity of the fluid.
We will see that the TI can already be suppressed by
sending back only a part of the battery current. 
We use the same geometry as before, and fix the inner diameter
to $d_i = 30\,$mm. We consider now a battery current 
$I_{bat}$ of 4 to 10\,kA, increase the
return current $I_{wire}$ step by step and measure the growth rate of
the TI (Fig.\  \ref{fig:rs_gr}). Evidently, for this realistic setting
the TI disappears already for return currents smaller then $I_{bat}$.
This fact might be utilized, e.g., in a stack of connected 
batteries in order to minimize the voltage drops in the central 
wires.

\subsection{Axial magnetic field}

It has been known for a long time that 
current-driven instabilities can be re-stabilized by applying 
an axial magnetic field. In plasma physics, this effect goes 
under the 
notion Kruskal-Shafranov condition for the so-called 
safety parameter, which is basically a ratio of axial to 
azimuthal magnetic field.
Related experiments
have shown that the kink-type instabilities can indeed be
switched off when a sufficiently strong axial 
magnetic field is applied \cite{Bergerson2006}.

\begin{figure}[h!]
\centerline{
\includegraphics[width=0.7\columnwidth]{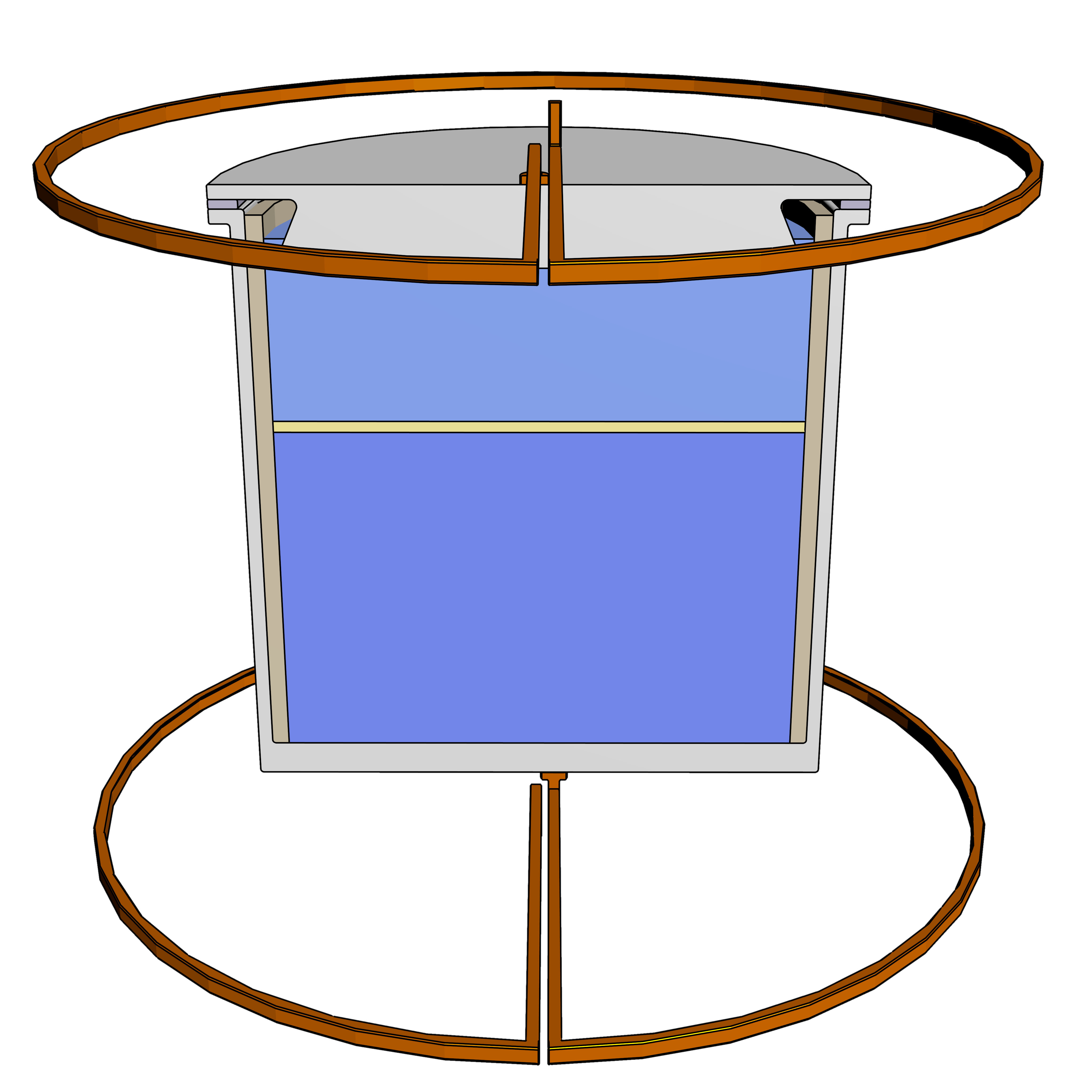}}
\caption{An axial magnetic field produced by a Helmholtz-like coil 
is an effective way to suppress the TI.}
\label{fig:bz_battery}
\end{figure}

In order to verify this effect for the battery case, 
we have again to take into account the finite 
height and the real viscosity and resistivity of the
materials. For the sake of concreteness, 
we assume a geometry as indicated in Fig.\  \ref{fig:bz_battery}, with 
a Helmholtz-like coil configuration to produce a rather homogeneous 
axial field in the battery. We will start with one single winding in
each part of the coil to see if this is already sufficient
for stabilization.

\begin{figure}[h!]
\centerline{
\includegraphics[width=\columnwidth]{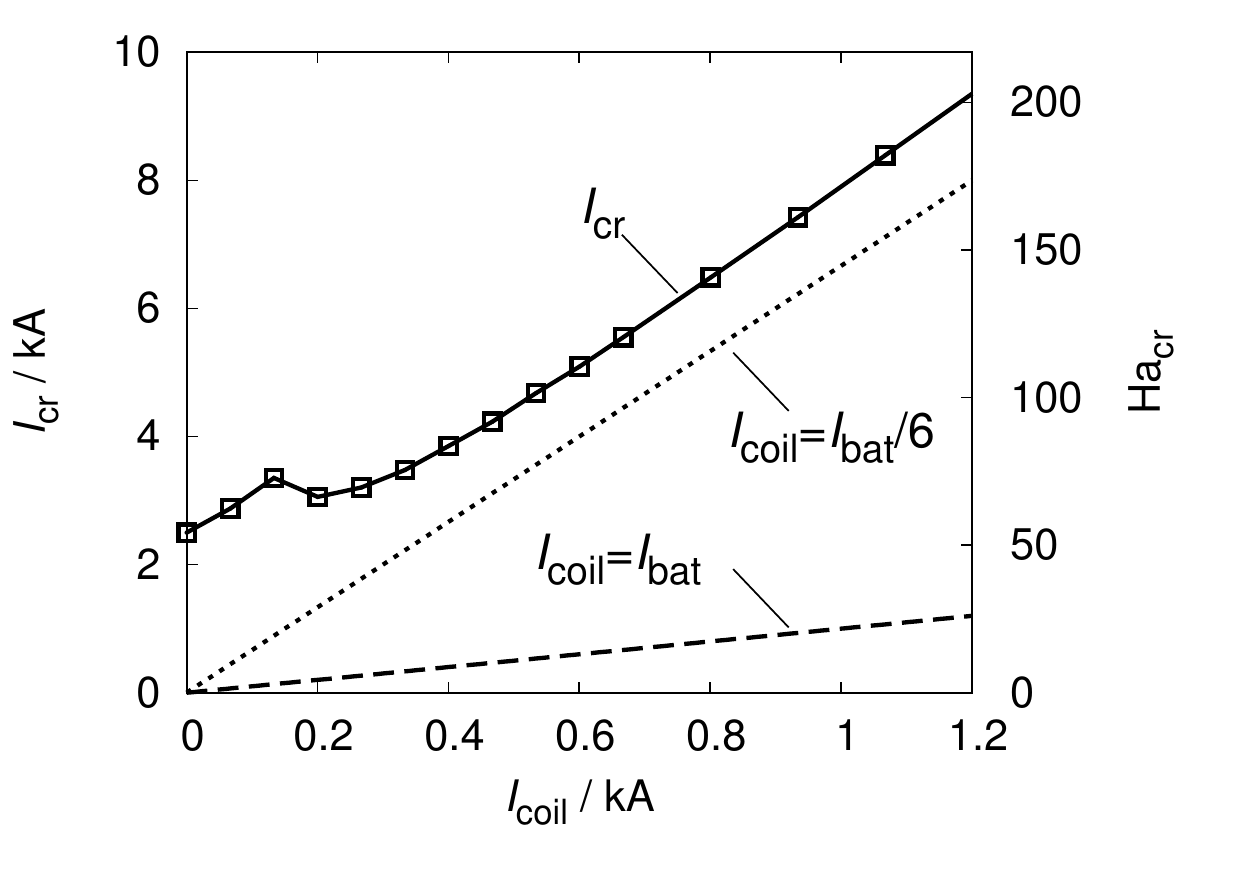}}
\caption{Critical current for onset of the TI in a cuboid Na electrode of the
  dimensions $1\times 1\times 0.5$\,m$^3$ with an external applied
magnetic field $B_z$, produced by a Helmholtz coil of a diameter of
$1.2$\,m.}
\label{fig:bz_Icr}
\end{figure}

For a cuboid geometry and a rather large Helmholtz coil,
the lower, dashed straight line in Fig.\  \ref{fig:bz_Icr} 
corresponds now to the
situation that the total battery current is also
used in the coil, i.e.\ $I_{bat}=I_{coil}$. 
Evidently, the critical current, as shown in the upper curve, 
increases with the coil current in such a way that there is always a
large margin between the actual battery current and the critical current. 
This means, in turn, that for 
suppressing the TI it would be sufficient to 
use only a part of the (dashed) battery current in the Helmholtz-like 
coil. This is indicated by the straight dotted curve in the middle,
which corresponds to the relation $I_{bat}=6 I_{coil}$. 
This is indeed of technical relevance since it allows to split
the current into a small part going through the coil and a larger
part going through an appropriate short-cut, 
thereby decreasing significantly the total voltage drop.

We note here that the configuration shown in Fig.\  
\ref{fig:bz_battery} is only a typical example. One can 
well imagine to use only one single winding, 
positioned approximately in the middle of the battery height.

\subsection{Horizontal magnetic field}
\begin{figure}[h!]
\centerline{
\includegraphics[width=0.7\columnwidth]{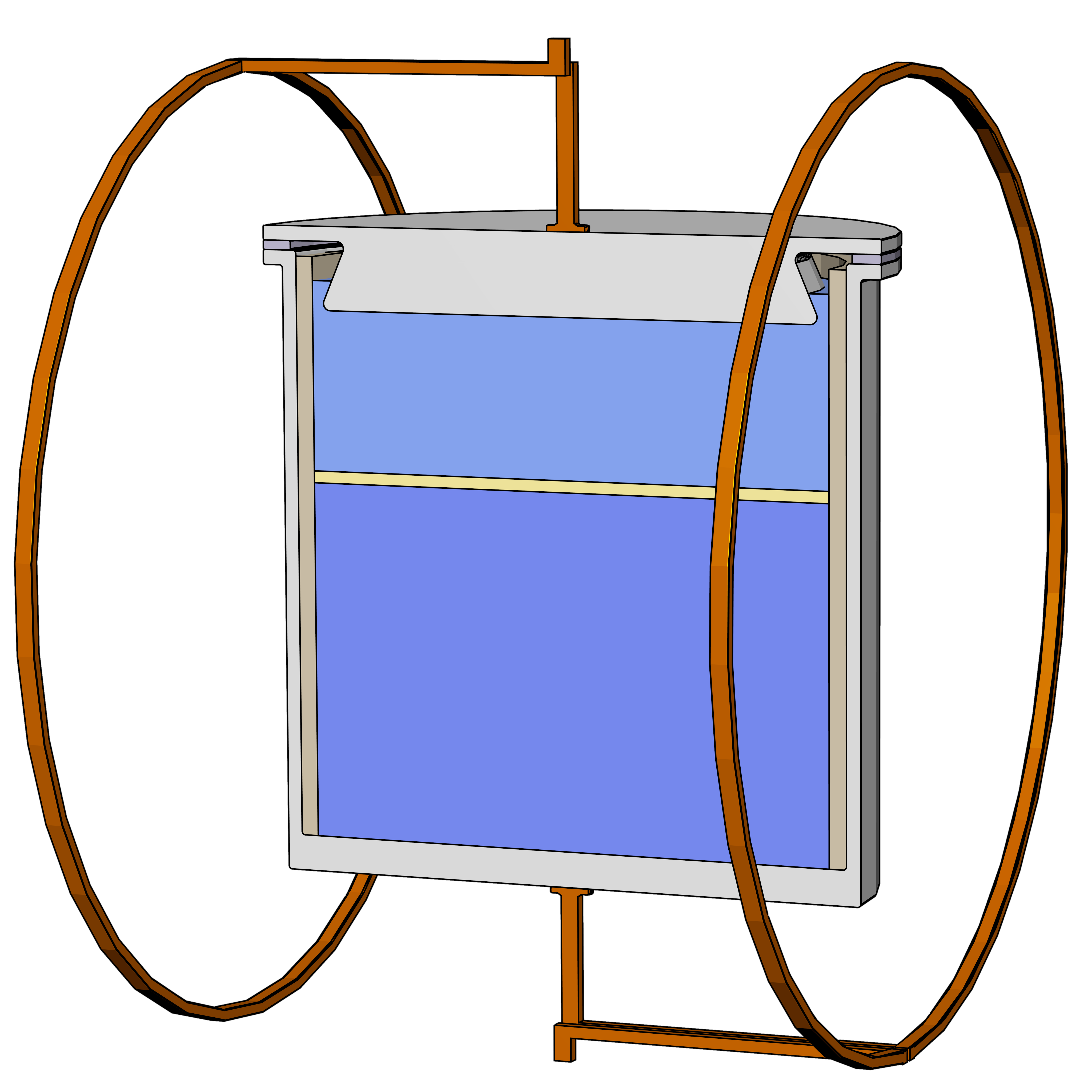}}
\caption{Example of a battery stabilization by a lateral Helmholtz-like 
coil producing a horizontal magnetic field.}
\label{fig:bx_battery}
\end{figure}

\begin{figure}[h!]
\centerline{
\includegraphics[width=\columnwidth]{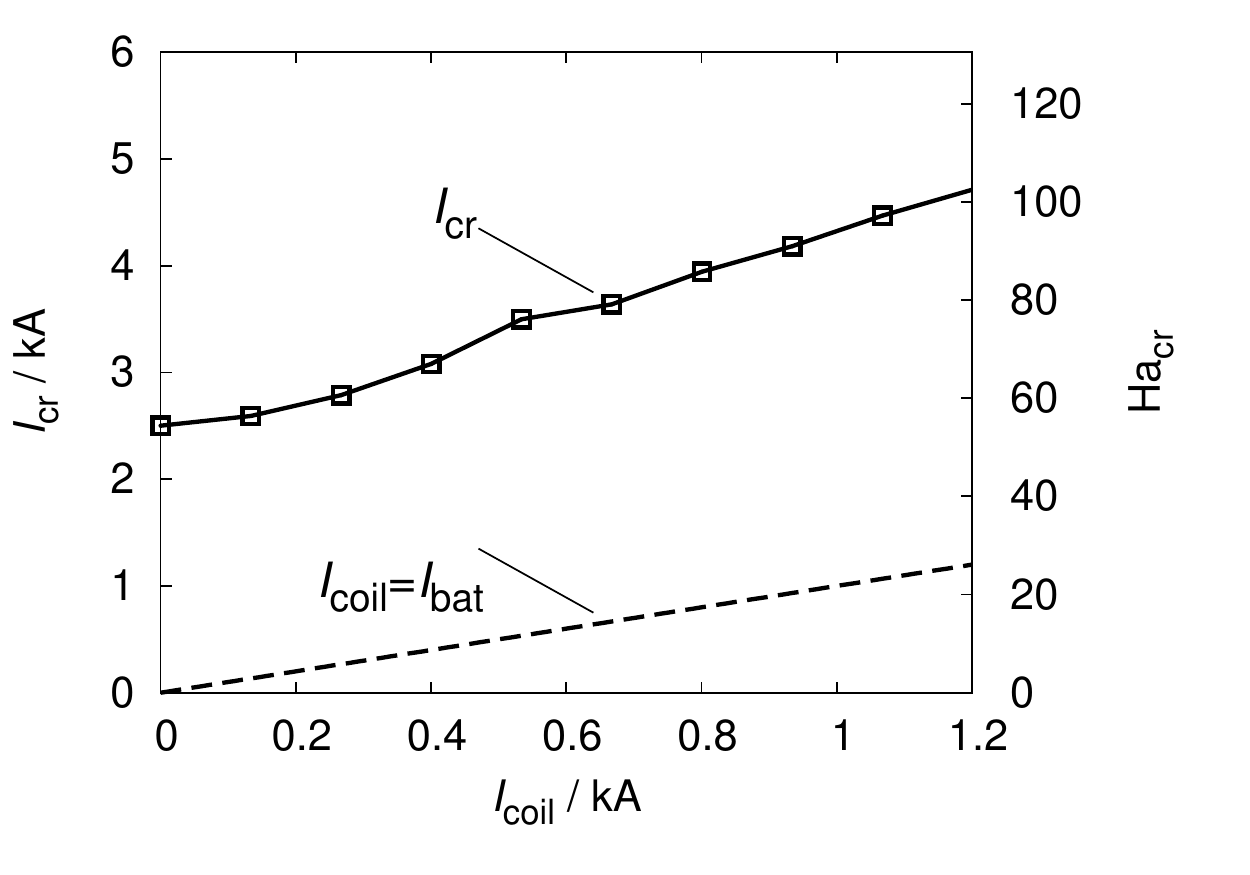}}
\caption{Critical current for onset of the TI in a cuboid Na electrode of the
  dimensions $1 \times 1 \times 0.5$\,m$^3$. The stabilizing horizontal magnetic
  field is produced by a Helmholtz coil with a diameter of $1.2$\,m.}
\label{fig:bx_Icr}
\end{figure}

If an axial magnetic field works so well in suppressing the TI,
we may ask if this functions also for an  applied
horizontal magnetic field 
(Fig.\  \ref{fig:bx_battery}).
Evidently, this is the case (see Fig.\  \ref{fig:bx_Icr}), 
although the critical current is not as steeply increasing with
the coil current as for the case of an axial field.
Yet, for certain stacked configurations of batteries
even this stabilization method might have advantages.

\section{Summary and outlook}
In this paper we have considered the 
susceptibility of large-scale liquid metal batteries to 
magnetohydrodynamic instabilities that are driven by the 
charging or discharging currents. 

Our main focus was on the kink-type Tayler instability 
that may become relevant for batteries with 
medium or large aspect ratios. 
First, we have characterized the dependence
of the TI on several material parameters. 
For this purpose we have utilized a new 
numerical code that allows to treat the TI for 
finite size geometries and for realistic material 
parameters of liquid metals. 
Not surprisingly, all the concrete parameter dependencies
turned out to collapse into a universal
dependence of the normalized growth rate 
and the Reynolds number on the Hartmann number.

In the main part of the paper, we have delineated five 
possible measures for suppressing the TI. In the first two methods 
we have varied the geometry of the liquid metal columns, 
either by reducing its height, 
or by inserting a central electrically insulating tube. 
The last three methods
utilized the fact that externally applied 
magnetic fields are able
to damp or even to suppress the TI. 
This has been exemplified by showing that both
a counter-current in the central bore as well as an axial or a 
horizontal magnetic field may re-stabilize the flow. In either 
case we have
indicated possible technical realizations, with some focus
on minimizing the material and/or the voltage drops in 
the necessary additional wires.

Having focused on the current driven TI,
we have to keep in mind the additional relevance of
interface instabilities which are well known
from aluminum production cells \cite{Davidson2000}.
The detailed investigation of those instabilities, and of 
their possible interaction with the TI, must be left 
for future work. 
This interaction aspect is highly non-trivial because 
measures that seem ideally suited for suppressing the TI 
(i.e.\ applying an axial field) might, in turn, lead 
to an enhancement of the interface instabilities. Numerical work 
on this interaction is presently in progress.

In a next step, we intend to investigate the influence of the TI
on the electrolyte layer experimentally. 
This is planned to be done in frame of the
DRESDYN project \cite{Dresdyn2012}, together with the 
magneto-rotational
instability, as recently investigated \cite{Stefani2009}.

\section*{Acknowledgment}
This work was supported by Helmholtz-Gemeinschaft Deutscher
Forschungszentren (HGF) in the frame of ``Initiative f\"ur mobile und
station\"are Energiespeichersysteme'' as well as in the frame of
Helmholtz Alliance ``Liquid metal technologies'' (LIMTECH). Fruitful
discussions with Marcus Gellert, J\={a}nis Priede, G\"unther R\"udiger, 
Martin Seilmayer, Steffen Landgraf and Andreas Bund on several aspects 
of the TI and liquid metal batteries are grateful acknowledged.












\end{document}